\documentclass[prd,twocolumn,preprintnumbers,superscriptaddress,nofootinbib]{revtex4}
\usepackage[a4paper, hdivide={1.91cm,,1.165cm}, vdivide={1.83cm,,3.6cm}]{geometry}

\usepackage{amstext,amssymb}
\usepackage{amsmath}
\usepackage{graphicx}
\usepackage[hyperfootnotes=false]{hyperref}
\usepackage{xspace}
\usepackage{color}
\usepackage{units}
\usepackage{slashed} 

\newcommand{\baz}{\begin{array}{cc}}
\newcommand{\bad}{\begin{array}{ccc}}
\newcommand{\ba}{\begin{array}{c}}
\newcommand{\ea}{\end{array}}
\newcommand{\be}{\begin{equation}}
\newcommand{\ee}{\end{equation}}
\newcommand{\bea}{\begin{eqnarray}}
\newcommand{\eea}{\end{eqnarray}}

\newcommand{\bi}{\begin{itemize}}
\newcommand{\ei}{\end{itemize}}
\newcommand{\bmt}{\begin{pmatrix}}
\newcommand{\emt}{\end{pmatrix}}
\newcommand{\bt}{\begin{tabular}}
\newcommand{\et}{\end{tabular}}

\newcommand{\benu}{\begin{enumerate}}
\newcommand{\eenu}{\end{enumerate}}

\definecolor{gesfred}{rgb}{1,0,0}

\definecolor{gesfblue}{rgb}{0.08,0.42,0.76}

\allowdisplaybreaks[4]

\begin{document}

\title{New physics effects on neutrinoless double beta decay from right-handed current}
\author{Shao-Feng Ge}
\email{gesf02@gmail.com}
\author{Manfred Lindner} 
\email{lindner@mpi-hd.mpg.de}
\affiliation{Max-Planck-Institut f\"ur Kernphysik, Saupfercheckweg 1, 69117 Heidelberg, Germany}
\author{Sudhanwa Patra}
\email{sudhanwa@mpi-hd.mpg.de}
\affiliation{Max-Planck-Institut f\"ur Kernphysik, Saupfercheckweg 1, 69117 Heidelberg, Germany}
\affiliation{Center of Excellence in Theoretical and Mathematical Sciences, \\
\hspace*{-0.2cm} 
Siksha 'O' Anusandhan University, Bhubaneswar-751030, India
}
\begin{abstract}
\noindent 
We study the impact of new physics contributions to neutrinoless double beta decay 
arising from right-handed current in comparison with the standard mechanism. If
the light neutrinos obtain their masses from Type-II seesaw within left-right symmetric model, where
the Type-I contribution is suppressed to negligible extent, the right-handed PMNS matrix is 
the same as its left-handed counterpart, making it highly predictable and can be tested
at next-generation experiments. It is very attractive, especially with recent cosmological 
constraint favoring
the normal hierarchy under which the neutrinoless double beta decay is too small to be
observed unless new physics appears as indicated by the recent diboson excess observed at
ATLAS. The relative contributions from left- and right-handed 
currents can be reconstructed with the ratio between lifetimes of two different
isotopes as well as the ratio of nuclear matrix elements. In this way, the theoretical
uncertainties in the calculation of nuclear matrix elements can be essentially avoided.
We also discuss the interplay of neutrinoless double beta decay measurements with 
cosmology, beta decay, and neutrino oscillation.
\end{abstract}

\pacs{98.80.Cq,14.60.Pq}
\maketitle

\noindent
\section{Introduction} 
Although the Standard Model (SM) of particle physics is extremely successful in explaining most 
of the experimental data up to current accelerator energies, it does not explain 
non-zero neutrino masses confirmed by solar, atmospheric and reactor oscillation data, 
as well as the mystery of dark matter and the matter dominance of present universe. 
These are the experimentally facts which motivate physics beyond the standard model 
and in addition there exist various theoretical reasons. 

Left-Right Symmetric Models (LRSM) \cite{Mohapatra:1974gc, Pati:1974yy, Senjanovic:1975rk,Senjanovic:1978ev,
Mohapatra:1979ia,Mohapatra:1980yp} are well motivated candidates 
of physics beyond SM. Among the reasons are: 
i) explaining light neutrino masses via seesaw mechanism with the natural embedding of right-handed neutrinos, 
ii) providing theoretical origin of maximal parity violation seen at weak interaction while conserved in strong 
and electromagnetic interaction,
iii) the fact that $B-L$ is a more attractive U(1) quantum number and the symmetry of U(1) charges of the fermions. 
With the scalar sector comprising of Higgs bidoublet and triplets, the light neutrino
mass is generated by type-I plus type-II seesaw mechanism. The Higgs scalars generate 
masses for light and heavy neutrinos which not only accommodates the Majorana nature of light and heavy 
neutrinos, but also leads to testable lepton number violation (LNV) at low energy experiments like 
neutrinoless double beta decay ($0\nu\beta\beta$) and at high energy experiments like the Large Hadron Collider (LHC, see refs. 
\cite{Tello:2010am} for a detailed discussions). 
The LHC can also probe right-handed current. Recently ATLAS~\cite{Aad:2015owa} claims to find a diboson excess 
which can  be explained by a 2\,TeV $W'$ via $pp \to W^\prime \to W\,Z$,
and CMS~\cite{Khachatryan:2014dka, Khachatryan:2014hpa,Khachatryan:2014gha} sets stringent lower bound
on the $W'$ mass through
$pp \to W^\prime \to \ell \ell+2 j$.
In the context of LRSM,
in principle, one can derive limits for right-handed neutrino mass
and mixing, $W_R$ 
mass and its mixing with the left-handed counterpart as well as the corresponding gauge 
coupling $g_R$.

On the other hand, neutrinoless double beta decay is a unique phenomena whose experimental 
observation would reveal the character of light neutrinos i.e, whether they have Majorana mass term~\cite{Majorana:1937vz} which violates lepton number.
If so, neutrinoless double beta decay can happen.
At present, the best lower limit on the decay half-life using 
$^{76}\mbox{Ge}$ is $T^{0\nu}_{1/2}\, > 2.1 \times 10^{25}$ yrs at 90\% C.L. 
from GERDA~\cite{Agostini:2013mzu} while the combined bound is $T^{0\nu}_{1/2}\, 
> 3.0 \times 10^{25}$ yrs. At the same time, the future experiment 
GERDA Phase-II~\cite{Smolnikov:2008fu} is expected to improve the half-life 
sensitivity to reach $T^{0\nu}_{1/2}\, > 2.0 \times 10^{26}$ yrs. 
For $^{136}\mbox{Xe}$, the derived lower limits on half-life from EXO-200 and KamLAND-Zen experiments are 
$T^{0\nu}_{1/2}\, > 1.6 \times 10^{25}$ yrs \cite{Gando:2012zm} 
and $T^{0\nu}_{1/2}\, > 1.9 \times 10^{26}$ yrs \cite{Auger:2012ar}, respectively. The combined 
limit from KamLAND-Zen Collaboration is $T^{0\nu}_{1/2}\, > 3.4 \times 10^{26}$ yrs 
at 90\% C.L.

Lepton number violating $0\nu\beta\beta$ transition could be induced by either light 
Majorana neutrinos or new physics. For the latter,
one has to go beyond SM and many models can contribute to neutrinoless 
double beta decay \cite{Mohapatra:1986su,Babu:1995vh,Hirsch:1995vr,Hirsch:1995ek,Hirsch:1996ye,
Racah:1937qq, Furry:1939qr,Tello:2010am, Das:2012ii,Barry:2013xxa,Dev:2014iva,Deppisch:2014zta,
Chakrabortty:2012mh, Awasthi:2013ff, Ge:2015bfa}. 
The nest-generation experiments of neutrinoless double decay can touch down to
the lower limit of inverted hierarchy (IH), corresponding to effective Majorana mass 
parameter $m^\nu_{ee}$ around $0.01\,\mbox{eV}$. Nevertheless, the latest cosmological 
bound on the mass sum of light active neutrinos \cite{Dell'Oro:2015tia} has indicated 
that the normal hierarchy (NH) is favored at $1 \sigma$ confidence level with tiny smallest 
mass. A direct consequence is that neutrinoless double beta decay 
is difficult to be observed at next-generation experiments if its only source
comes from the standard mechanism. If the result of observation turns out to
be the opposite, namely neutrinoless double beta decay is observed but cannot 
be explained by the standard mechanism, it clearly indicates physics beyond SM.

With these motivations, both hints from  the diboson excess observed at ATLAS
for heavy $W'$ around $2\,\mbox{TeV}$ and cosmological preference for NH with tiny mass 
scale and invisible neutrinoless double beta decay even at next-generation experiments, 
we examine the effect of Type-II seesaw contributions 
to neutrinoless double beta decay arising from purely right-handed current within 
LRSM. The predicted effect mass can saturate the current experimental bounds
making it testable in the future even with NH being favored.

The paper is organized as follows: After briefly summarizing the 
standard mechanism of neutrinoless double beta decay in Sec.~\ref{sec:review} we consider 
new physics effect from purely right-handed current in Sec.~\ref{sec:new}. Next we 
compare the new physics contribution with the standard mechanism and derive lower 
limit on the absolute scale of light neutrinos from current experimental limit in Sec.~\ref{sec:comparison}.
In Sec.~\ref{sec:ratio}, we use half-life and
nuclear matrix element (NME) ratios to reconstruct the relative contributions 
of the standard mechanism and new physics to effective Majorana mass parameter. This approach to 
distinguishing different sources has a significant advantage that
the theoretical uncertainties in the calculation of NME can be essentially avoided.
We conclude our paper in Sec.~\ref{sec:conclusion}.

\section{Neutrinoless double beta decay}
\label{sec:review}
\begin{figure}[!h]
\includegraphics[scale=0.58]{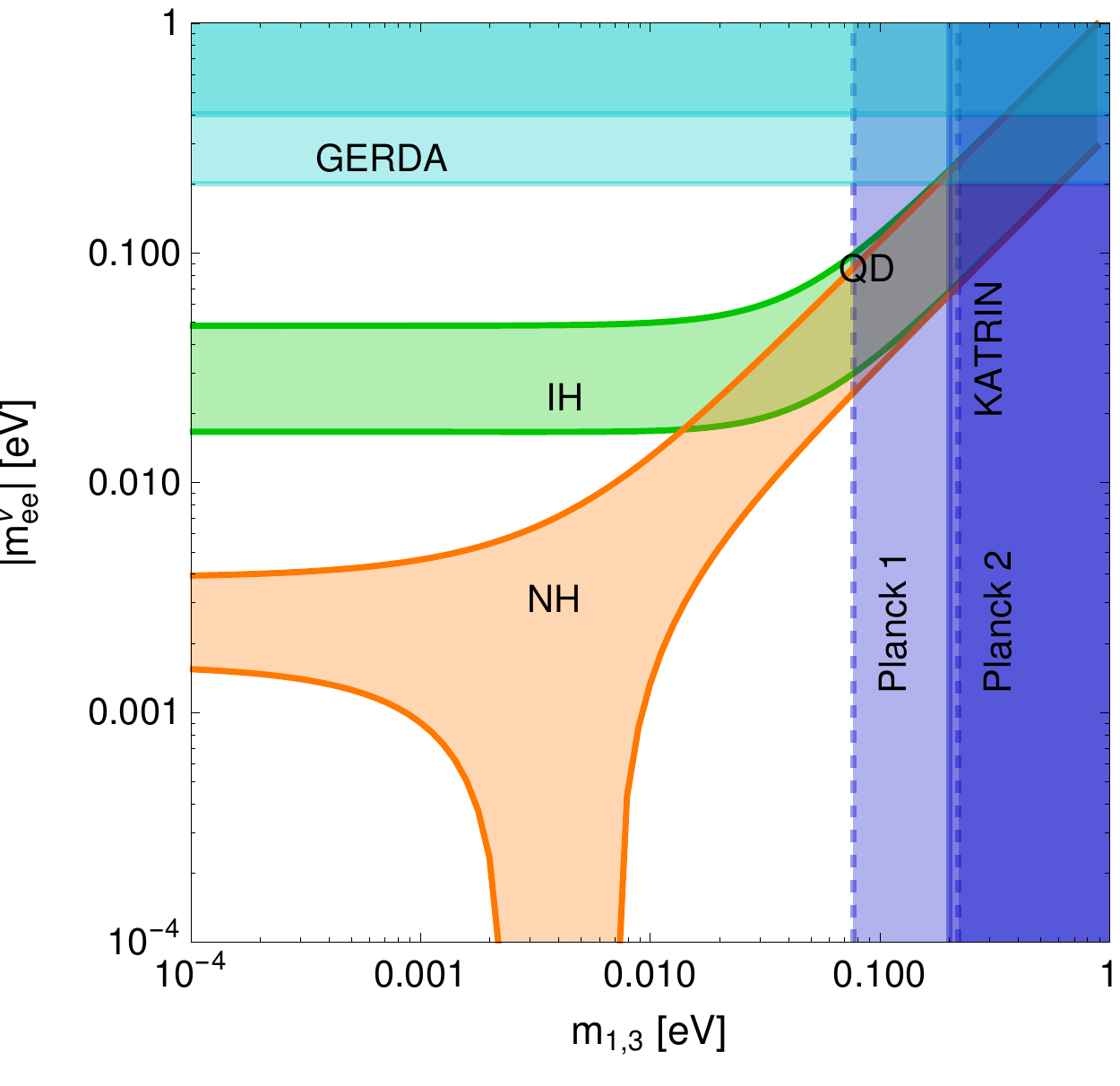}
\caption{Effective Majorana mass for neutrinoless double beta decay within the 
         standard mechanism as a function of the lightest neutrino mass, $m_1$ 
         for NH (yellow band) and $m_3$ for IH (green band). The oscillation parameters are
         varied within the $3\sigma$ ranges taken from ref.~\cite{Forero:2014bxa} while 
         the Majorana CP phases $\alpha$ and $\beta$ are treated as free parameters.
         The horizontal band are excluded regions of the effective Majorana mass parameter
         $m^\nu_{ee}$ by current neutrinoless double
         beta decay measurement (GERDA), while the vertical bands are excluded regions
         of the mass scale by beta decay (KARTRIN) and cosmological observations (Planck).}
\label{plot:onubb-std}
\end{figure}

Observing neutrinoless double beta decay can not only reveal LNV
but also provide crucial information about the absolute scale of light neutrino masses and 
the associated mass-generation mechanism. If light Majorana neutrinos are the only  
source of $0\nu\beta\beta$ transition, we can express the half-life as
\begin{align}
\frac{1}{T_{1/2}^{0\nu}} &=
	G^{0\nu}_{01}\left| {\cal M}^{0\nu}_\nu\right|^2 \left| \eta_\nu \right|^2 
	=G^{0\nu}_{01} \left| \frac{{\cal M}^{0\nu}_\nu}{m_e}\right|^2   
	\left| m^\nu_{ee} \right|^2\, ,
\label{eq:half-life_std}
\end{align}
which contains three important factors: 
i) the NME ${\cal M}^{0\nu}_\nu$, 
ii) the phase space factor $G^{0\nu}_{01}$, and 
iii) a dimensionless particle physics parameter -- a measure of lepton number violation $\eta_\nu \equiv m^\nu_{ee} / m_e$ which is a function of
     neutrino masses, mixing angles, and CP phases,
\begin{subequations}
\begin{eqnarray}
\left| m^\nu_{ee} \right|^2&=&
  \bigg| |U^2_{e1}|\, m_1 + |U^2_{e2}|\, m_2 e^{i \alpha} + |U^2_{e3}|\, m_3 e^{i \beta} \bigg| \qquad \\ &=&
  \left| c^2_s c^2_r m_1 + s^2_s c^2_r m_2 e^{i \alpha} + s^2_r m_3 e^{i \beta} \right|^2 \,,
\label{eq:mee-std}
\end{eqnarray}
\end{subequations}
where $m_i$ are the mass eigenvalues of light neutrinos, $\alpha, \beta$ are the two Majorana 
CP phases, and $U_{ei}$ are the elements of PMNS mixing matrix. We present the variation of 
$\left| m^\nu_{ee} \right|$, due to the unknown Majorana CP phases, as a function of the lightest 
neutrino mass ($m_1$ for NH and $m_3$ for IH) in Fig.\ref{plot:onubb-std}. 
The notation of mixing angles is defined as $(c_\alpha, s_\alpha)\equiv (\cos\theta_\alpha, \sin\theta_\alpha)$ where 
$\theta_a\equiv \theta_{23}$ is the atmospheric mixing angle, $\theta_r\equiv \theta_{13}$ 
the reactor mixing angle, and $\theta_s\equiv \theta_{12}$ the solar mixing angle. 
In addition, the oscillation data are sensitive to two mass-squared differences,
the solar mass-squared difference $\Delta m^2_{\rm s} 
\equiv m^2_2-m^2_1$ and the atmospheric mass-squared difference $\Delta m^2_{\rm {a}}\equiv |m^2_3-m^2_1|$. Since we know only the magnitude of the atmospheric
mass-squared difference with its sign unknown, there exit two different mass patterns 
among light neutrinos: NH with $m_1 < m_2 < m_3$ and IH with $m_3 < m_1 < m_2$. 
The $3\sigma$ global fit ranges for the oscillation parameters are summarized
in Table~\ref{table-osc}.

The effective Majorana mass parameter of neutrinoless 
double beta decay that predicted by the standard mechanism is displayed in Fig.~\ref{plot:onubb-std} with light-red (green) band 
for NH (IH),
respectively. The horizontal bands are bounds from neutrinoless double beta decay measurements
which can provide crucial information about the absolute scale of light neutrino mass
and Majorana CP phases. In addition, the mass scale can also be bounded from 
i) the measurement of beta decay parameter $m_\beta=\sqrt{\sum_i U^2_{ei} m_i}$ for which the existing 
bound is $m_\beta < 0.2$\,eV from KATRIN \cite{Osipowicz:2001sq}, 
ii) The limits on the effective Majorana mass parameter from
various isotopes \cite{Ge:2015bfa}, 
\begin{eqnarray*}
&&\left| m^\nu_{ee} \right| \leq (0.22 \sim 0.53)\,\mbox{eV} \quad \quad \mbox{For}~^{76}\mbox{Ge}\,, \nonumber \\
&&\left| m^\nu_{ee} \right| \leq (0.36 \sim 0.90)\,\mbox{eV} \quad \quad \mbox{For}~^{100}\mbox{Mo}\,, \nonumber \\
&&\left| m^\nu_{ee} \right| \leq (0.27 \sim 1.00)\,\mbox{eV} \quad \quad \mbox{For}~^{130}\mbox{Te}\,, \nonumber \\
&&\left| m^\nu_{ee} \right| \leq (0.15 \sim 0.35)\,\mbox{eV} \quad \quad \mbox{For}~^{136}\mbox{Xe}\,.
\end{eqnarray*}
iii) Cosmological observations which provide constraints on the sum of light neutrino masses $m_\Sigma\equiv \sum_i m_i$. They have been shown as
vertical bands in Fig.~\ref{plot:onubb-std}. Note that these constraints have only
touched down to the quasi-degenerate region, $m_1\simeq m_2 \simeq m_3$ and $m_{\rm lightest} > 
\sqrt{\Delta m^2_{\rm {a}}}$, with the beta-decay and cosmological constraints approximately at $\sum_i m_i/3 \simeq m_{\rm lightest}\simeq m_\beta 
\geq \left| m^\nu_{ee} \right|$ which is the same for both hierarchies. The current bound on the 
sum of light neutrino mass is $m_\Sigma < 0.23$\,eV derived from Planck+WP+highL+BAO data 
(Planck1) at 95\% C.L. while $m_\Sigma < 1.08$\,eV from Planck+WP+highL (Planck2) at 95\% C.L. 
\cite{Ade:2013zuv}. 
It is claimed that the latest cosmological constraint can be approximated by
gaussian distribution, $m_\Sigma = 22 \pm 62 \, \mbox{meV}$ \cite{Dell'Oro:2015tia}. At $1\sigma$ C.L., the
mass sum is smaller than $84\,\mbox{meV}$ which is below the smallest value for IH.
In this sense, the NH is favored.
There is no hope
to probe the normal hierarchy within the standard mechanism, even for next-generation experiments. 
On the contrary, if there is new physics contribution in addition to the standard
mechanism, the neutrinoless double beta decay can be sizable. In the following part,
we discuss one possibility of LRSM with Type-II seesaw dominance which is fully
predictable and can be tested at next-generation experiments.
\begin{table}[!h]
\begin{tabular}{cc}
        \hline 
Oscillation Parameters  &  3$\sigma$ range       \\
        \hline \hline
$\Delta m^2_{\rm {s}} [10^{-5}\,\mbox{eV}^2]$              & 7.11-8.18     \\
$|\Delta m^2_{\rm {a}}| [10^{-3}\,\mbox{eV}^2]$ (NH)       & 2.30-2.65     \\
$|\Delta m^2_{\rm {a}}| [10^{-3}\,\mbox{eV}^2]$ (IH)       & 2.20-2.54     \\
\hline
$\sin^2\theta_{s}$                                        & 0.278-0.375     \\
$\sin^2\theta_{r}$ (NH)                                   & 0.0177-0.0294   \\
$\sin^2\theta_{r}$ (IH)                                   & 0.0183-0.0297   \\
        \hline
\end{tabular}
\caption{The $3\sigma$~values of the global fit oscillation parameters, 
          the mass-squared differences and mixing angles~\cite{Forero:2014bxa}.}
\label{table-osc}
\end{table}

\begin{widetext}
\begin{figure*}[t]
\includegraphics[scale=0.5]{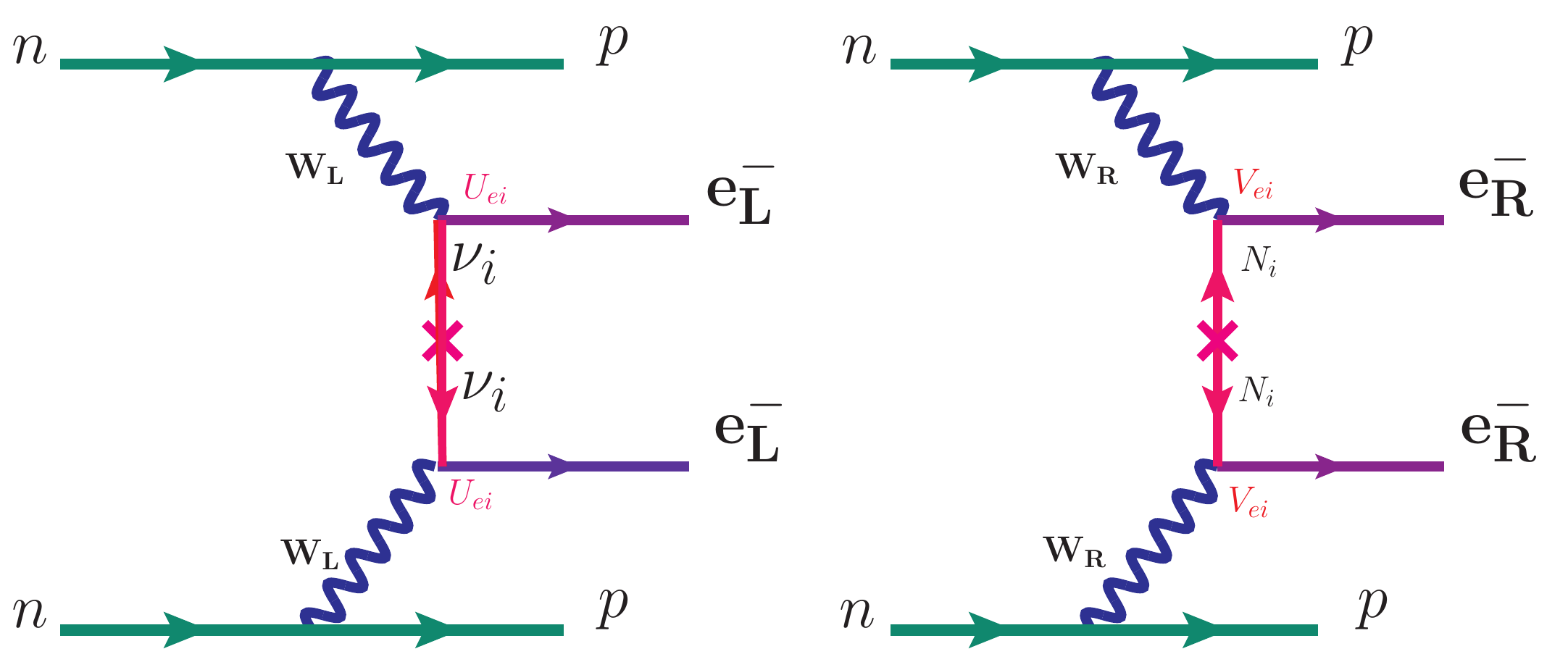}
\caption{Feynman diagrams of neutrinoless double beta decay processes. The left-panel 
         is for purely left-handed current contribution by the standard mechanism while the right-panel is for purely 
         right-handed current contribution by LRSM with Type-II 
         seesaw dominance.}
\label{feyn:onubb-LR}
\end{figure*}
\end{widetext}

\section{Effect of New Physics Contribution}
\label{sec:new}
In this section we briefly discuss a simple left-right symmetric model 
as the source of new physics that can contribute to neutrinoless double beta decay
and is fully predictable.
The left-right symmetric model \cite{Mohapatra:1974gc, Pati:1974yy, Senjanovic:1975rk,
Senjanovic:1978ev,Mohapatra:1979ia,Mohapatra:1980yp} is based 
on the gauge group $SU(2)_L \times SU(2)_R \times U(1)_{B-L}$--omitting $SU(3)_C$ 
structure for simplicity--where $SU(2)_L$ interchanges with $SU(2)_R$ under
parity. The quarks and leptons individually transform as
\begin{eqnarray}
  Q_{L}
=
\left\lgroup
\begin{matrix}
  u_{L} \\
  d_{L}
\end{matrix}
\right\rgroup
\equiv 
  \left[ 2,1, \frac 1 3 \right] 
& , & 
  Q_{R}
=
\left\lgroup
\begin{matrix}
  u_{R} \\
  d_{R}
\end{matrix}
\right\rgroup
\equiv
  \left[ 1,2,{\frac{1}{3}} \right] \,,
\nonumber 
\\
  \ell_{L}
=
\left\lgroup
\begin{matrix}
  \nu_{L}\\
  e_{L}
\end{matrix}
\right\rgroup
\equiv
  [2,1,-1] 
& , & 
  \ell_{R}
=
\left\lgroup
\begin{matrix}
  \nu_{R} \\
  e_{R}
\end{matrix}
\right\rgroup
\equiv
  [1,2,-1] \,. 
\nonumber
\end{eqnarray}
Spontaneous symmetry breaking is implemented as usual with the help of triplet
and bidoublet Higgs fields,
\begin{eqnarray}
  \Delta_{L,R} 
\equiv 
\left\lgroup
\begin{matrix} 
  \delta_{L,R}^+/\sqrt{2} & \delta_{L,R}^{++} \\ 
  \delta_{L,R}^0 & -\delta_{L,R}^+/\sqrt{2} 
\end{matrix}
\right\rgroup, 
\quad
  \Phi 
\equiv 
\left\lgroup
\begin{matrix} 
  \phi_1^0 & \phi_2^+ \\ 
  \phi_1^- & \phi_2^0 
\end{matrix}
\right\rgroup,
\end{eqnarray}
with quantum numbers $(3,1,2)$, $(1,3,2)$, and $(2,2,0)$, respectively. 
The bidoublet connects left- and right-handed fermions to provide Dirac mass term
while the triplet provides Majorana mass term to left- or right-handed neutrinos.
After symmetry breaking, the neutrino mass matrix can be written as,
\begin{equation}
	M_\nu
= 
\left\lgroup
\begin{matrix}
		M_L  & m_D   \\
   	m^T_D & M_R
\end{matrix}
\right\rgroup \,,
\label{eqn:numatrix}       
\end{equation}
where, $M_L= f_L \langle \Delta_L \rangle = f v_L $ ($M_R= f_R \langle \Delta_R \rangle = f v_R$)
is the LH (RH) Majorana masses for light (heavy) neutrinos while $m_D= y v_1 + \tilde{y} v_2$ 
is the Dirac neutrino mass term connecting light and heavy neutrinos. 
The parity is first broken by $v_R$, the vacuum expectation value (VEV) of 
the right-handed triplet $\Delta_R$, reducing $SU(2)_L \times SU(2)_R \times U(1)_Y$ to the SM gauge 
structure $SU(2)_L \times U(1)_Y$ which is further broken by
the VEV of the bidoublet $\Phi$. This symmetry breaking sequence naturally requires 
$v_R$ to be much larger than the electroweak scale. On the other hand, the VEV $v_R$ of 
the left-handed triple $\Delta_L$ is small. With natural Yukawa couplings, the mass
matrix elements of $M_\nu$ have a hierarchical structure,
$M_R \gg m_D \gg M_L$. The light neutrino mass matrix can be obtained via Type-I+II
seesaw mechanism,
\begin{equation}
	m_\nu = M_L - m_D  M^{-1}_R m^T_D
	= m_\nu^{II} + m_\nu^I \,.
\label{neutrino-mass}
\end{equation}

The Majorana nature of the light and heavy neutrinos, arising from the presence of the scalar triplets $\Delta_{L,R}$
with charge $B-L=2$, implies that they can mediate neutrinoless double beta decay \cite{Tello:2010am, 
Das:2012ii,Barry:2013xxa,Dev:2014iva,Deppisch:2014zta,Chakrabortty:2012mh, 
Awasthi:2013ff}, as shown in Fig.~\ref{feyn:onubb-LR}. The left- and right-handed
currents contribute independently without interference. The other diagrams contributing 
to neutrinoless double beta decay via doubly charged scalar triplet exchange, $\lambda$-diagram 
involving $W_L-W_R$ mediation and $\eta$-diagram are suppressed, for 
details, see refs.\cite{Chakrabortty:2012mh, Awasthi:2013ff, Barry:2013xxa}.

\subsection{Type-II seesaw dominance}

As shown in (\ref{neutrino-mass}), the light neutrino mass matrix receives
two independent contributions from Type-I and Type-II seesaw mechanisms. Although
the left- and right-handed Yukawa/mass matrices are connected with each other, 
$M_L \propto M_R$, due to parity, the Dirac mass matrix $m_D$ is independent of 
the others.
Experimentally, we have measured the oscillation parameters of the light neutrinos. 
The information is not enough to constrain both $M_L(M_R)$ and $m_D$. While the 
light neutrino mass matrix $m_\nu$ can be reconstructed, the heavy one $M_R$
is out of our knowledge. On the other hand, both the light and heavy neutrino mass
matrices are involved in the neutrinoless double beta decay process as shown in
Fig.~\ref{feyn:onubb-LR}. It is difficult to make a specific prediction unless we 
can somehow know the heavy neutrino mass matrix $M_R$. Fortunately, Type-II seesaw 
dominance can save the situation here \cite{Tello:2010am,Deppisch:2014zta,Chakrabortty:2012mh,Dev:2013vxa,Barry:2013xxa}.

With Type-II seesaw dominance, which can be realized by suppressing 
the Dirac mass term $m_D$, the mass matrices of the left- and right-handed neutrinos
are proportional to each other,
\begin{equation}
  m_\nu
=
  M_L
\propto
  M_R \,.
\end{equation}
A direct consequence is that the left- and right-handed neutrinos share the
same mixing matrix,
\begin{equation}
 V^{PMNS}_R=
  V^{PMNS}_L \,,
\label{eq:equality-VLR}
\end{equation}
in the diagonal basis of charged leptons.
The unmeasured right-handed neutrino mixing can then be fully reconstructed.

Actually, the equality between the left- and right-handed PMNS mixing matrices
can also come from the lepton sector. To see this let us first write down the Yukawa 
couplings that are responsible for the lepton mass matrix,
\begin{equation}
  \overline L_L (Y_\ell \Phi + Y'_\ell \Phi^\dagger) L_R
+ h.c. \,,
\end{equation}
where $\Phi^\dagger \equiv \sigma_2 \Phi^* \sigma_2$ is the CP mirror of the bidoublet
$\Phi$ with $\tau_2$ being the second Pauli matrix. When the Higgs bidoublet
develops a non-zero VEV, 
\begin{equation}
  \Phi
=
\left\lgroup
\begin{matrix}
  v_1 e^{i \theta_1} \\
& v_2 e^{i \theta_2}
\end{matrix}
\right\rgroup \,,
\end{equation}
the leptons receive a mass matrix,
\begin{equation}
  M_\ell
=
  Y_\ell v_2 e^{i \theta_2} 
+ Y'_\ell v_1 e^{- i \theta_1} \,.
\end{equation}
If the Higgs doublet fails to develop trivial CP phases or if 
the two Yukawa matrices $Y_\ell$ and $Y'_\ell$ commute with 
each other, the lepton mass matrix can commute with its Hermitian conjugate,
\begin{equation}
  M_\ell M^\dagger_\ell
=
  M^\dagger_\ell M_\ell \,.
\end{equation}
A direct consequence is that the left- and right-handed leptons share the same mixing matrix,
\begin{equation}
  V_{\ell,L}
=
  V_{\ell,R} \,.
\end{equation}
If the neutrino mixing is trivial, $V_\nu = I$, then the PMNS mixing matrices come 
solely from the charged lepton mixing, reproducing
(\ref{eq:equality-VLR}).

For all cases, the mass eigenvalues 
for light as well as heavy neutrinos are related as,
\begin{equation}
	M_i \propto m_i \,,
\label{eq:typeII_dom_eigen_rel:a}
\end{equation}
where $M_i$ and $m_i$ are physical masses for light active and heavy Majorana neutrinos. 
Fixing $M_{\rm max}$( $M_3$ for NH and $M_2$ for IH), the mass relation can be expressed 
as,
\begin{subequations}
\begin{eqnarray}
	M_{i} &= \frac{m_i}{m_{3}} M_{3}, \text{ for NH}, 
\label{eq:NH_massrel-typeII} \\
	M_{i} &= \frac{m_i}{m_{2}} M_{2}, \text{ for IH}.
\label{eq:IH_massrel-typeII} 
\end{eqnarray}
\end{subequations}

\section{Comparison between New Physics and Standard Mechanism}
\label{sec:comparison}
The inverse of half-life for a particular isotope, with
new physics effects realized via a 
left-right symmetric model assuming Type-II seesaw dominance, 
contributes to neutrinoless double beta decay 
therefore in the following way,
\begin{subequations}
\begin{eqnarray}
\frac{1}{T_{1/2}^{0\nu}}&=&
	G^{0\nu}_{01} \bigg[\left|\mathcal{M}_\nu^{0\nu}\cdot \eta_\nu \right|^2  
	        +\left| \mathcal{M}_N^{0\nu} \cdot \eta_N \right|^2 \bigg] 
	\\
	&=&
 	G^{0\nu}_{01} \left| \frac{\mathcal{M}_\nu^{0\nu}}{m_e} \right|^2 \left| m^{\left(\nu+N\right)}_{\rm ee} \right|^2 ,
\end{eqnarray}
\label{eq:half-life_typeII}
\end{subequations}
where $\left| m^{\left(\nu+N\right)}_{\rm ee} \right|^2 \equiv \left| m^\nu_{\rm ee} \right|^2+ \left| m^N_{\rm ee} \right|^2$ 
with explicit analytic expressions under Type-II seesaw dominance, 
\begin{subequations}
\begin{eqnarray}
\hspace{-5mm}
&&
  \left| m^\nu_{\rm ee} \right|
\equiv
  \left| c^2_s c^2_r m_1 + s^2_s c^2_r m_2 e^{i \alpha} + s^2_r m_3 e^{i \beta} \right| \,,
\label{eq:mee-std} \\
\hspace{-5mm}
&&
  \left| m^N_{\rm ee} \right|_{NH}
\equiv
  \frac{C_N}{M_{3}}  \bigg|c^2_s c^2_r \frac{m_3}{m_1} 
       +  s^2_s c^2_r \frac{m_3}{m_2} \,e^{i \alpha} + s^2_r\,e^{i \beta} \bigg| \,,
\label{eq:NH_mee-typeII} 
 \\
\hspace{-5mm}
&&
  \left| m^N_{\rm ee} \right|_{IH} 
\equiv
  \frac{C_N}{M_2} \left|c^2_s c^2_r \frac{m_2}{m_1} 
+ s^2_s c^2_r e^{i \alpha} +\frac{m_2}{m_3}  s^2_r e^{i \beta} \right|  \,.
\qquad
\label{eq:IH_mee-typeII}
\end{eqnarray}
\label{eq:mee}
\end{subequations}
The coefficient $C_N \equiv \langle p^2 \rangle  \left(M_{W_L}/M_{W_R}\right)^4 (g_R / g_L)^4$ contains the typical momentum transfer $\langle p \rangle \approx 100 \mbox{MeV}$.
Here, the lepton number violating particle 
physics parameters are $\eta_\nu$ and $\eta_N (m^\nu_{ee}, m^{N}_{ee})$, while $G^{0\nu}_{01}$ is the phase space factor, 
and $\mathcal{M}_\nu^{0\nu}$($\mathcal{M}_N^{0\nu}$) the NMEs derived for virtual light and heavy particle 
exchanged diagram, respectively, whose values can be found in Table~\ref{tab:nucl-matrix}.
\begin{table}
\centering
\vspace{10pt}
\begin{tabular}{lccc}
\hline
Isotope & $G^{0\nu}_{01}\,\left({\rm yr.}^{-1}\right)$ & ${\cal M}^{0\nu}_\nu$ & 
${\cal M}^{0\nu}_N$  \\
\hline 
$^{76}$Ge   & $5.77 \times 10^{-15}$ & 2.58--6.64 & 233--412  \\ 
$^{136}$Xe  & $3.56 \times 10^{-14}$ & 1.57--3.85 & 164--172  \\
\hline
\end{tabular}
\caption{The Phase space factor $G^{0\nu}_{01}\,\left({\rm yr.}^{-1}\right)$ and 
          nuclear matrix elements (NMEs) for two different isotopes $^{76}$Ge and 
          $^{136}$Xe \cite{Meroni:2012qf}.}
\label{tab:nucl-matrix}
\end{table}
The $SU(2)_L$ and $SU(2)_R$ coupling constants $g_L$ and $g_R$, respectively, need not to 
be the same for LRSM.
In addition to an overall factor $C_N/M_3$ for NH or $C_N/M_2$ for IH, $m^N_{ee}$ share the same set 
of input parameters as $m^\nu_{ee}$.
To estimate the lepton number violating parameter arising from right-handed current,
It is necessary to know oscillation parameters, the Majorana CP phases, and the mass scales.
\begin{itemize}
\item {\bf Oscillation parameters:} 
Neutrino oscillation experiments has measured the two mass-squared differences and the three
mixing angles. Nevertheless only two mixing angles, the reactor mixing angle $\theta_r$ and the solar
mixing angle $\theta_s$, are relevant in the effective Majorana mass parameter. The error of these four
oscillation parameters can introduce uncertainty in the prediction of $|m^\nu_{ee}|$, especially
the solar mixing angle $\theta_s$ \cite{Dueck:2011hu}. With the current global fit \cite{Forero:2014bxa},
the upper limit on the minimal half-life for IH varies by a factor of $2.5$ at $3 \sigma$ C.L.
Fortunately, the same set of oscillation parameters is shared by reactor neutrino experiments
and hence can be precisely measured by next-generation medium-baseline experiments like JUNO 
\cite{An:2015jdp}. The uncertainty from oscillation parameters can be reduced to below 1\% which
is essentially negligible \cite{Ge:2015bfa}.

\item {\bf CP Phases:}
Although the Dirac CP phase $\delta_D$ also enters the effective Majorana mass parameter $|m^\nu_{ee}|$ and
$|m^N_{ee}|$, it will not manifest itself in the predicted values. This is because it is always 
associated with one of the two Majorana CP phases $\alpha$ and $\beta$ which are completely unknown. 
For simplicity, the Dirac CP phase has been omitted in the expressions. The variation of $\alpha$
and $\beta$ renders the predicted values of $|m^\nu_{ee}|$ and $|m^N_{ee}|$ to span in a wide band.
In Appendix \ref{app:geometry}, we generalize the geometrical picture of obtaining the minimal 
and maximal values within standard mechanism to LRSM-Type II seesaw dominance.

\item {\bf Mass Scales:}
In addition to oscillation parameters and CP phases, the mass scales of light and heavy neutrinos
are still unknown yet. For light neutrinos, we fix the mass scale by hand since there is no
measurement at all. On the other hand, the contribution from heavy neutrinos involve two mass scales,
the heavy neutrino and $W_R$ masses.
First, we need to fix the heavy neutrino mass scales by fixing $M_3$ for NH or $M_2$ for IH.
Second, the $W_R$ mass also enters through $C_N$. When combined, the net effect is an overall factor,
$1/(M^4_{W_R} M_3)$ for NH or $1/(M^4_{W_R} M_2)$ for IH. A natural realization is $M_{2,3}$ should be roughly
of the same scale as $M_{W_R}$. In other words, the low-energy neutrino phenomenology is highly related 
with collider signature. If the di-boson excess \cite{Aad:2015owa} 
observed at ATLAS really comes from LRSM with mass $M_{W_R} \approx 2\,\mbox{TeV}$, its contribution
to low energy neutrino phenomenology is large enough to be observed by next-generation neutrinoless 
double beta decay experiments.
\end{itemize}

\begin{widetext}
\begin{figure*}[t]
\includegraphics{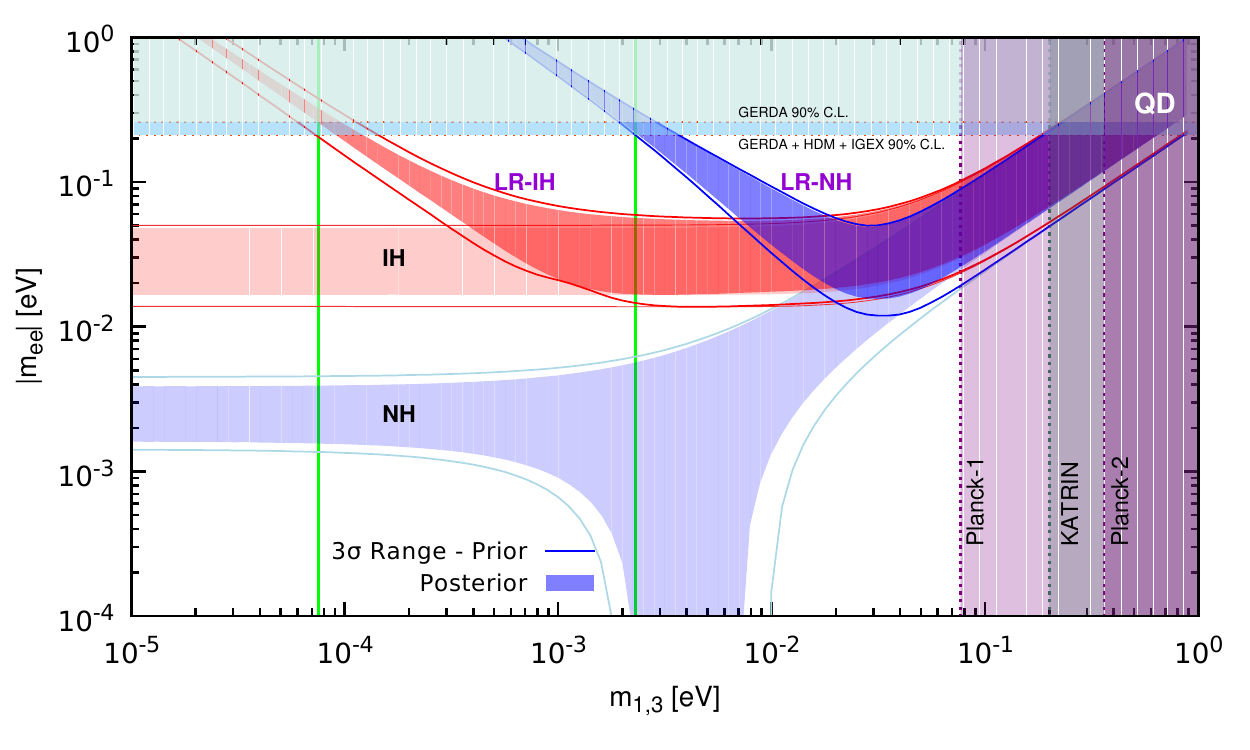}
\caption{The SM and LRSM-Type II contributions to the effective Majorana mass parameter
         $|m_{ee}|$ as a function of the lightest neutrino mass, $m_1$ for NH (dark blue band) 
         and $m_3$ for IH (dark red band), within the $3 \sigma$ ranges of both prior from 
         current global fit (band) and posterior after including JUNO (enveloping curve). 
         The horizontal bands are the current experimental bounds on the effective mass of
         neutrinoless double beta decay, from which lower limits (vertical green lines on the left) 
         on the lightest neutrino mass can be extracted. The vertical bands on the right are beta decay
         and cosmological constrains excluding most part of quasi-degenerate region.}
\label{plot:meeLR_comparision}
\end{figure*}
\end{widetext}
We show in Fig.~\ref{plot:meeLR_comparision}
the lepton number violating effective 
Majorana mass parameters vs lightest neutrino mass for $^{76}\mbox{Ge}$ which is used by GERDA~\cite{Agostini:2013mzu}. For comparison,
we show both the total contribution and the one from standard 
mechanism without new physics. 
The numerical analysis has been performed using $3\sigma$ prior and posterior 
distributions from current global fit and the one improved with JUNO \cite{An:2015jdp}.
As studied in \cite{Ge:2015bfa} the combination of short- and medium-baseline
reactor experiments can achieve precise measurement of the oscillation
parameters involved in neutrinoless double beta decay. Here, we adopt the same configuration
of JUNO with 53\,km of baseline, 20\,kt of detector, 36\,GW of thermal power, 6 years of running
(300 effective days for each year), and energy resolution $3\%/\sqrt{E/\mbox{MeV}}$. This roughly
reproduce the results of the JUNO yellow book \cite{An:2015jdp}. It is clearly shown in
Fig.~\ref{plot:meeLR_comparision} that the uncertainty
from oscillation parameters has significant effect on the prediction of the effective Majorana mass 
parameter within not only the standard mechanism but also the total contribution.

In Fig.\ref{plot:meeLR_comparision} we have fixed $M_{W_R} \simeq 2$\,TeV, 
$g_R \approx 2/3 g_L$, and $M_N=1$\,TeV which can satisfy both ATLAS \cite{Aad:2015owa}
and CMS \cite{Khachatryan:2014dka} excesses.
With this typical mass assignment indicated by collider signature, the contribution from new physics 
is not small and actually can dominate over the standard mechanism in low mass region. In the 
quasi-degenerate region, there is no big difference between the total contribution and the 
standard mechaism. As the mass scale of light neutrinos decreases, the new physics contribution
keeps growing and stops the total one from decreasing to zero for both hierarchies. The minimal value
is within the sensitivity region of next-generation neutrinoless double beta decay experiments which
can touch down to around $0.01\,\mbox{eV}$. When the smallest mass eigenvalue $m_{1,3}$ of the light
neutrinos further decreases, the new physics contribution dominates and can finally saturate the 
current experimental bound placed by GERDA \cite{Agostini:2013mzu}, 
Heidelberg-Moscow \cite{KlapdorKleingrothaus:2000sn}, 
and IGEX \cite{Aalseth:2002rf}. Since no experiment has observed neutrinoless
double beta decay, we can derive lower limit on the
absolute scale of light neutrinos, 
\begin{subequations}
\begin{eqnarray}
  m_1 > 2.3 \, \mbox{meV} & \mbox{for} & \mbox{NH} \,,
\\
  m_3 > 0.075 \, \mbox{meV} & \mbox{for} & \mbox{IH} \,.
\end{eqnarray}
\end{subequations}
as indicated by solid vertical lines in Fig.~\ref{plot:meeLR_comparision}. 
This is totally different from the standard mechanism based on which neutrinoless double beta
decay experiments can only provide upper limits. For IH, the standard mechanism cannot provide
lower limit which may be possible for NH. On the contrary, the LRSM-Type II contribution can
place both lower and upper limits on the mass scale of light neutrinos. 

\begin{widetext}
\begin{figure*}[!t]
\includegraphics{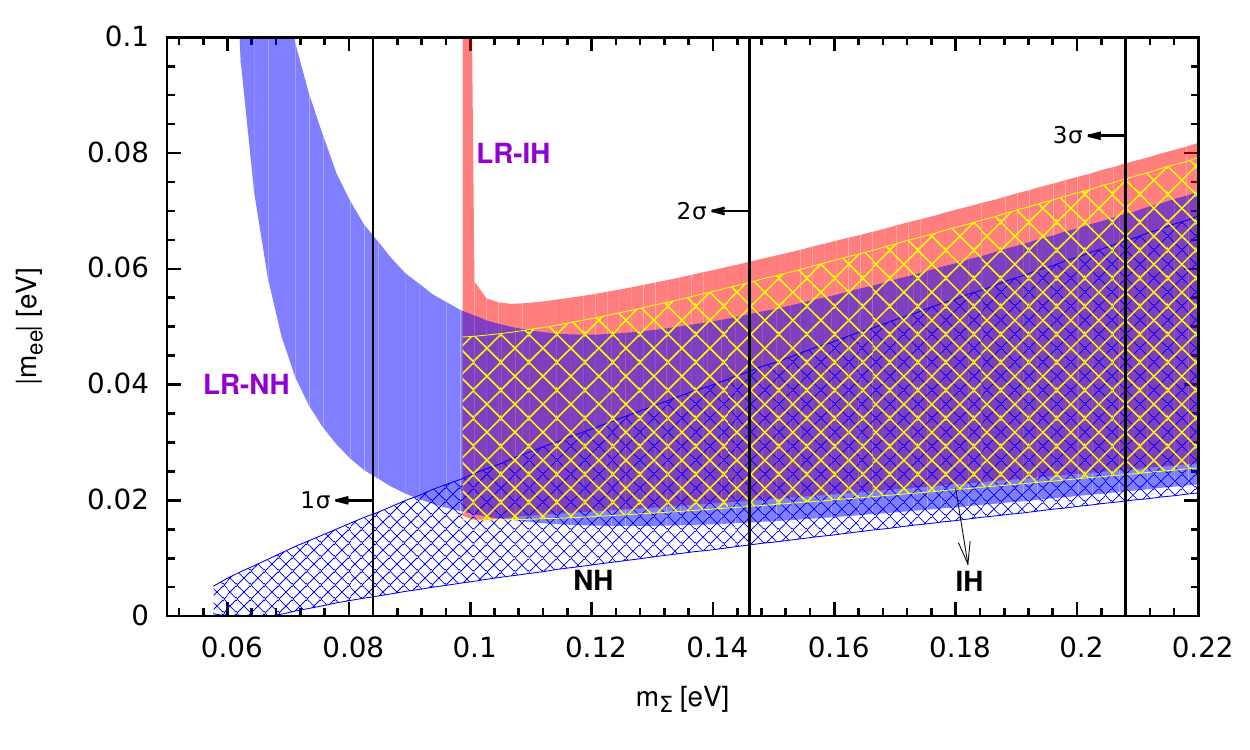}
\caption{Allowed region of $|m_{ee}|$ as a function of the light neutrino mass sum $m_\Sigma$. 
         The standard mechanism contribution is plotted with hatched bands, the dense blue one
         for NH and sparse yellow one for IH, while the new physics contribution with shaded bands,
         the blue one for LR-NH and the red one for LR-IH.
         The comparison between SM and LRSM-Type II seesaw contributions to the effective Majorana 
         mass parameter $|m_{ee}|$ as a function of the mass sum has been presented using JUNO 
         data and bound from cosmology. The limits on mass sum at 1$\sigma$, 2$\sigma$ 
         and 3$\sigma$ C.L., $m_\Sigma < 84 \, \mbox{meV}$, $m_\Sigma < 146 \, \mbox{meV}$, 
         and $m_\Sigma < 208 \, \mbox{meV}$, respectively, are shown as vertical lines.}
\label{plot:meemsum_comparision}
\end{figure*}
\end{widetext}

The lower limits obtained from current neutrinoless double beta decay measurements
are consistent with the cosmological constraint on the sum of 
light neutrino masses \cite{Seljak:2004xh,Costanzi:2014tna, Palanque-Delabrouille:2014jca},
\begin{eqnarray}
&&m_\Sigma < \mbox{84\,meV}\quad \quad ~(1\sigma~\mbox{C.L.}) \,, \nonumber \\
&&m_\Sigma < \mbox{146\,meV}\quad \quad (2\sigma~\mbox{C.L.}) \,,           \\
&&m_\Sigma < \mbox{208\,meV}\quad \quad (3\sigma~\mbox{C.L.}) \,. \nonumber
\end{eqnarray}
For clarity, we replot the effective Majorana mass parameter in Fig.~\ref{plot:meemsum_comparision}
as a fuction of the mass sum. It clearly shows that at more than $1 \sigma$ C.L., NH is
favored over IH. If so, the low-energy neutrinoless double beta decay cannot be observed 
even at next-generation
experiments, if no new physics is present. Fortunately, the diboson excess observed at ATLAS provides
a timely way out.

\section{Separating the SM and LRSM Contributions}
\label{sec:ratio}

\begin{figure*}[!t]
\includegraphics[width=0.48\textwidth,height=6.5cm]{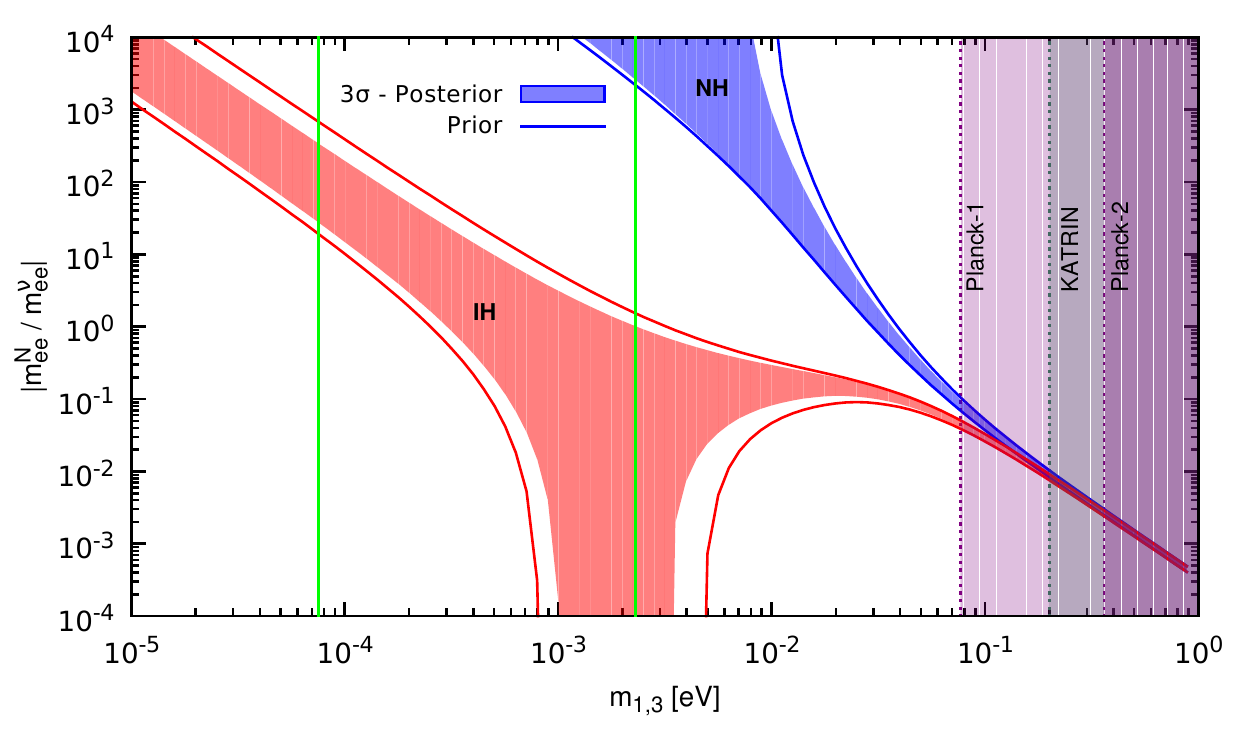}
\includegraphics[width=0.48\textwidth,height=6.5cm]{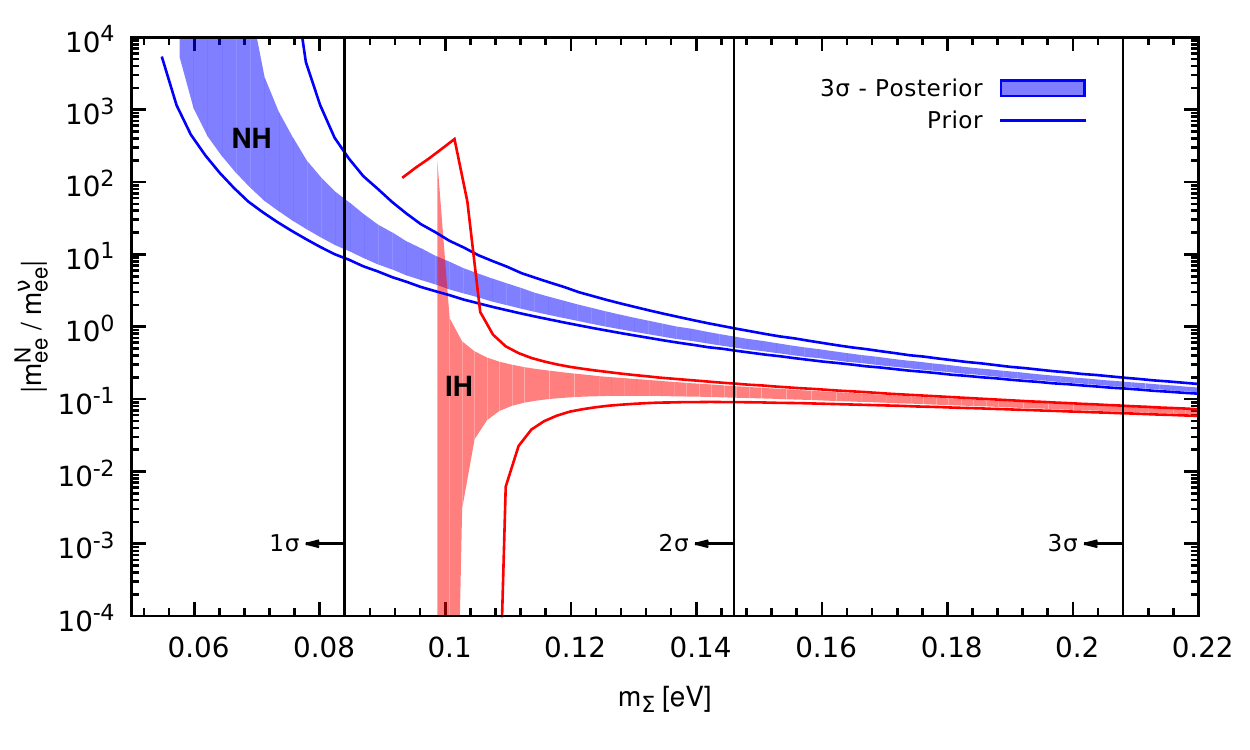}
\caption{Ratio between effective Majorana mass parameters $|m^N_{ee}/m^\nu_{ee}|$ as a function of 
         lightest neutrino mass, $m_1$ for NH and $m_3$ for IH in the left. The vertical green lines
	       comes from the saturation limit in Fig.~\ref{plot:meeLR_comparision}. In the right, we show
         the ratio $|m^N_{ee}/m^\nu_{ee}|$ as a function of the mass sum $m_\Sigma$ together with 
         recent cosmological constraints. Here $|m^\nu_{ee}|$ is the standard mechanism 
         contribution with $W_L$--$W_L$ mediation while $|m^N_{ee}|$ the 
         LRSM-Type-II seesaw contribution via $W_R$--$W_R$ mediation.}
\label{fig:meeRatio}
\end{figure*}

In LRSM with Type-II seesaw, the two contributions are linearly combined as shown
in (\ref{eq:half-life_typeII}). Using two isotopes, $(A_i, Z_i)$ and $(A_j, Z_j)$, 
it is possible to separate the SM and non-interfering new physics contributions 
\cite{Faessler:2011qw,Meroni:2012qf},
\begin{subequations}
\begin{eqnarray}
  |m^\nu_{ee}|^2
& = &
  \frac {|M^{0 \nu}_{N,j}|^2/T_i G_i - |M^{0 \nu}_{N,i}|^2/T_j G_j}
				{|M^{0 \nu}_{\nu,i}|^2 |M^{0 \nu}_{N,j}|^2 - |M^{0 \nu}_{N,i}|^2 |M^{0 \nu}_{\nu,j}|^2} \,,
\\
  |m^N_{ee}|^2
& = &
  \frac {|M^{0 \nu}_{\nu,i}|^2/T_j G_j - |M^{0 \nu}_{\nu,j}|^2/T_i G_i}
				{|M^{0 \nu}_{\nu,i}|^2 |M^{0 \nu}_{N,j}|^2 - |M^{0 \nu}_{N,i}|^2 |M^{0 \nu}_{\nu,j}|^2} \,.
\qquad
\end{eqnarray}
\end{subequations}
However, there is 
sizable uncertainty in the theoretical calculation of the nuclear matrix elements
which implies that the extracted effective Majorana mass parameters 
$|m^\nu_{ee}|$ and $|m^N_{ee}|$ have uncertainties. 

Fortunately, the theoretical uncertainties can be significantly reduced by taking the
ratios between nuclear matrix elements as well as between half-lifes of different isotopes.
These ratios can be used to obtain the ratio
of new physics and standard mechanism effective Majorana mass parameters,
\begin{equation}
  \left| \frac {m^N_{ee}}{m^\nu_{ee}} \right|^2
=
  \left| \frac {M^{0 \nu}_{\nu,j}}{M^{0 \nu}_{N,i}} \right|^2
  \frac {T_i G_i}{T_j G_j}
  \frac {\left| \dfrac{M^{0 \nu}_{\nu,i}}{M^{0 \nu}_{\nu,j}} \right|^2 - \dfrac{T_j G_j}{T_i G_i}}
        {\left| \dfrac{M^{0 \nu}_{N,j}}{M^{0 \nu}_{N,i}} \right|^2  - \dfrac{T_i G_i}{T_j G_j}} \,.
\label{eq:ratio}
\end{equation}
One can therefore expect that this ratio will be more precisely determined than the individual effective
Majorana mass parameters.

In Fig.~\ref{fig:meeRatio}, we show the ratio $|m^N_{ee}/m^\nu_{ee}|$ as a function of
the lightest neutrino mass in the left panel. Its value can
span several orders of magnitude. In the quasi-degenerate region,
the ratio is typically below $0.1$. For NH, it grows with decreasing mass scale. On the other
hand it can experience a short range, $1 \, \mbox{meV} \lesssim m_3 \lesssim 4 \, \mbox{meV}$, 
of touching down to the bottom, although the maximal value keeps growing with decreasing
$m_3$. Across the whole range, the ratio for NH is always larger than its value for IH.
Together with the lower limit on the lightest mass extracted from current neutrinoless 
double beta decay measurements, as depicted in Fig.~\ref{plot:meeLR_comparision}, the ratio
is constrained to be smaller than $10^3$ for IH while it can go well above $10^4$ for NH.
If neutrinoless double beta decay are observed with two isotopes at future
experiments to reconstruct the ratio $|m^N_{ee}/m^\nu_{ee}|$, it is possible to determine 
the neutrino mass hierarchy.

According to the definition (\ref{eq:mee}) of effective Majorana mass parameters,
the contribution of the right-handed sector actually can be separated from light neutrino parameters
since it only appears as an overall factor $C_N/M_3$ for NH or $C_N/M_2$ for IH.
From the measured ratio (\ref{eq:ratio}) we can reconstruct,
\begin{subequations}
\begin{eqnarray}
  \frac {\left| c^2_s c^2_r m_1 + s^2_s c^2_r m_2 e^{i \alpha} + s^2_r m_3 e^{i \beta} \right|}
				{\bigg|c^2_s c^2_r \frac{m_3}{m_1} +  s^2_s c^2_r \frac{m_3}{m_2} \,e^{i \alpha} + s^2_r\,e^{i \beta} \bigg|}
& = &
  \left| \frac {m^N_{ee}}{m^\nu_{ee}} \right|
  \frac {M_3}{C_N} \,,
\\
  \frac {\left| c^2_s c^2_r m_1 + s^2_s c^2_r m_2 e^{i \alpha} + s^2_r m_3 e^{i \beta} \right|}
				{\left|c^2_s c^2_r \frac{m_2}{m_1} + s^2_s c^2_r e^{i \alpha} +\frac{m_2}{m_3}  s^2_r e^{i \beta} \right|}
& = &
  \left| \frac {m^N_{ee}}{m^\nu_{ee}} \right|
  \frac {M_2}{C_N} \,, \qquad
\end{eqnarray}
\end{subequations}
which can be compared with collider measurements on the right-handed sector parameters contained
in $M_{2,3}/C_N$. This can be used to add another input
to determine the Majorana CP phases. The degeneracy between $\alpha$ and $\beta$ can then be
eliminated.

\section{Conclusions}
\label{sec:conclusion}
We have shown that new physics contributions to neutrinoless double beta decay induced 
by right-handed current can saturate the present experimental bound.
Then, the neutrinoless double beta decay experiment is still possible to see 
a signal for NH which together with small mass scale is favored by the recent bound on 
$m_\Sigma$ from cosmology. In comparison, if only the standard mechanism
contributes, it is difficult to see a signal of neutrinoless double beta decay even
at next-generation experiments. Fortunately, collider signature from ATLAS and CMS
has indicated heavy gauge boson $W'$ around $2\,\mbox{TeV}$, pointing to a timely way out.

We studied a simple framework of LRSM to provide the new 
physics effects to neutrinoless double beta decay. The neutrino mass mechanism is 
governed by Type-II seesaw dominance such that the 
mass eigenvalues and mixing matrices of light and heavy neutrinos are correlated 
with each other. To illustrate the idea, we carried out numerical estimation for 
$M_{W_R} \approx 2 \, \mbox{TeV}$ in order to show the typical distribution.

We derived the corresponding lower limits on the absolute scale of light neutrinos, 
$m_1 \gtrsim 2.3 \, \mbox{meV}$ for NH and $m_3 \gtrsim 0.075 \, \mbox{meV}$ for IH by saturating 
the limits set by current neutrinoless double beta decay experiments. Different from the
standard mechanism, where only upper limit on the mass scale can be extracted from neutrinoless
double beta decay measurements, LRSM-Type II seesaw allows both upper and lower limits. 
In addition, the total effective Majorana mass parameter is bounded from below and well within
the sensitivity reach of the next-generation experiments.
To distinguish the two non-interfering contributions from the light and heavy neutrinos,
we take ratios between the half-lifes and nuclear matrix elements for two isotopes. In this
way, the theoretical uncertainties in the calculation of nuclear matrix elements can be avoided. 
Once measured and LRSM-Type II seesaw is confirmed, the ratio $|m^N_{ee}/m^\nu_{ee}|$ can be
interpreted to distinguish the light neutrino mass hierarchy. Further, supplemented with
collider measurements of the right-handed sector parameters, the two Majorana CP phases
$\alpha$ and $\beta$ can be uniquely determined without degeneracy.

\section*{Acknowledgements}
The work of SP is partially supported by the Department of Science and 
Technology, Govt.\ of India under the financial grant SB/S2/HEP-011/2013 
and by the Max Planck Society in the project MANITOP.  

\appendix
\section{Maximum and Minimum values for Effective Majorana Mass parameter}
\label{app:geometry}

Given mass eigenvalues and mixing angles, the effective
Majorana mass parameter can still vary due to the two unknown Majorana CP phases.
A geometric picture \cite{Xing:2014yka} has been developed to obtain the maximal 
and minimal values analytically. This can still apply when Type-II seesaw 
contribution from LRSM is also included.

As a result of (\ref{eq:equality-VLR}) within Type-II seesaw dominance, there 
are two contributions to the effective Majorana mass parameter,
\begin{subequations}
\begin{eqnarray}
&&
  \left| m^\nu_{ee} \right|^2
=
  \left| c^2_s c^2_r m_1 + s^2_s c^2_r m_2 e^{i \alpha} + s^2_r m_3 e^{i \beta} \right|^2 \,,
\\
&&
  \left| m^N_{ee} \right|^2
=
  \left| c^2_s c^2_r \widetilde m_1 + s^2_s c^2_r \widetilde m_2 e^{i \alpha} + s^2_r \widetilde m_3 e^{i \beta} \right|^2 \,,
\qquad
\label{eq:MeeR1}
\end{eqnarray}
\end{subequations}
with concrete form of $\widetilde m_i$ to be found in the text.
The interesting feature is that, in Type-II seesaw, the same set of mixing parameters,
including the two Majorana CP phases $\alpha$ and $\beta$, the two mixing angles
$\theta_r$ and $\theta_a$, as well as the Dirac CP phase $\delta_D$, are shared 
between the two contributions. It provides
a possibility of using the same geometrical picture to get the minimal and
maximal values of $|m_{ee}|$ due to the variation of $\alpha$
and $\beta$. For convenience, let us redefine $|m^\nu_{ee}|$ and $|m^N_{ee}|$ as,
\begin{subequations}
\begin{eqnarray}
&&
  \left| m^\nu_{ee} \right|^2
\equiv
  |f_1 + f_2 e^{i \alpha} + f_3 e^{i \beta}|^2 \,,
\\
&&
  \left| m^N_{ee} \right|^2
\equiv
  |F_1 + F_2 e^{i \alpha} + F_3 e^{i \beta}|^2 \,.
\end{eqnarray}
\end{subequations}
When expanded, they can be written in terms of real functions,
\begin{subequations}
\begin{eqnarray}
  \left| m^\nu_{ee} \right|^2
& = &
  \sum_i f^2_i
+ 2 f_1 f_2 \cos \alpha
\nonumber
\\
& + &
  2 f_1 f_3 \cos \beta
+ 2 f_2 f_3 \cos (\alpha - \beta) \,,
\\
  \left| m^N_{ee} \right|^2
& = &
  \sum_i F^2_i
+ 2 F_1 F_2 \cos \alpha
\nonumber
\\
& + &
  2 F_1 F_3 \cos \beta
+ 2 F_2 F_3 \cos (\alpha - \beta) \,.
\end{eqnarray}
\end{subequations}
Both contributions share the same functional form of $\alpha$ and $\beta$.
After combination, the functional form remains,
\begin{eqnarray}
	\left| m_{ee} \right|^2
& = &
  \sum_i (f^2_i + F^2_i)
+ 2 G_1 G_2 \cos \alpha
\nonumber
\\
& + &
  2 G_1 G_3 \cos \beta
+ 2 G_2 G_3 \cos (\alpha - \beta) \,,
\end{eqnarray}
with,
\begin{subequations}
\begin{eqnarray}
  G_1 G_2
& \equiv &
  f_1 f_2 + F_1 F_2 \,,
\\
  G_1 G_3
& \equiv &
  f_1 f_3 + F_1 F_3 \,,
\\
  G_2 G_3
& \equiv &
  f_2 f_3 + F_2 F_3 \,.
\end{eqnarray}
\end{subequations}
The redefined elements $G_i$ can be readily solved,
\begin{subequations}
\begin{eqnarray}
  G_1
& = &
  \sqrt{\frac {(f_1 f_2 + F_1 F_2)(f_1 f_3 + F_1 F_3)}{f_2 f_3 + F_2 F_3}} \,,
\\
  G_2
& = &
  \sqrt{\frac {(f_1 f_2 + F_1 F_2)(f_2 f_3 + F_2 F_3)}{f_1 f_3 + F_1 F_3}} \,,
\\
  G_3
& = &
  \sqrt{\frac {(f_1 f_3 + F_1 F_3)(f_2 f_3 + F_2 F_3)}{f_1 f_2 + F_1 F_2}} \,.
\end{eqnarray}
\end{subequations}
The combined effective electron mass can be formulated as,
\begin{equation}
  \left| m_{ee} \right|^2
\hspace{-1mm}
=
\hspace{-1mm}
  \sum_i (f^2_i + F^2_i - G^2_i)
+ |G_1 + G_2 e^{i \alpha} + G_3 e^{i \beta}|^2 \,.
\label{eq:MeeTot}
\end{equation}
Then, the geometrical picture of finding the minimal and maximal values 
by varying $\alpha$ and $\beta$ as elaborated in \cite{Xing:2014yka} can readily
apply to the last term in (\ref{eq:MeeTot}).

\bibliographystyle{utcaps_mod}
\bibliography{onubb_LR}
\end{document}